\documentclass[prl,twocolumn,amsmath,amssymb,groupedaddress,superscriptaddress]{revtex4-1}
\usepackage{graphicx} \usepackage{bm}
\usepackage{epsf} \usepackage{epsfig}
\usepackage{graphics}

\def\be{\begin{equation}}

\def\ee{\end{equation}}

\def\Ttr{\textmd{Tr}}

\def\Rre{\textmd{Re}}

\def\be{\begin{equation}}
\def\ee{\end{equation}}
\def\bea{\begin{eqnarray}}
\def\eea{\end{eqnarray}}

\begin{document}

\title{On the structure of the exact master equation}

\author{In{\'e}s \surname{de Vega}}
\affiliation{Department of Physics and Arnold Sommerfeld Center for Theoretical Physics, Ludwig-Maximilians-Universit{\"a}t M{\"u}nchen, Theresienstr. 37, 80333 Munich, Germany}
\email{ines.vega@physik.uni-muenchen.de}

\begin{abstract}
We derive a master equation from the exact stochastic Liouville von-Neumann (SLN) equation \cite{stockburger2002,wallsbook}. The latter depends on two correlated noises and describes exactly the dynamics of an oscillator (which can be either harmonic or present an anharmonicity) coupled to an environment at thermal equilibrium. The newly derived master equation is obtained by performing analytically the average over different noise trajectories. It is found to have a complex hierarchical structure that might be helpful to explain the convergence problems occurring when performing numerically the stochastic average of trajectories given by the SLN equation \cite{koch2008,koch2010}. 
\end{abstract}

\date{\today}

\maketitle
\section{Introduction}
In the past decades, there has been an increasing interest in analyzing the dynamics of open quantum systems (OQS) coupled to an environment. Much effort has been devoted to the derivation of master equations that describe the dynamics of the reduced density matrix of the system \cite{breuerbook}. 

An exact master equation can be derived for the case in which the OQS is a harmonic oscillator (or a collection of them) \cite{hu1992,halliwell1996}. However, when the OQS contains an anharmonicity, giving rise for instance to dynamics restricted to a two state basis, no master equation has been derived up to now without making some approximation. 
Whereas different master equations have been derived within the so-called weak coupling approximation, i.e. assuming that the coupling between the system and the environment is small compared to other energy scales of the problem, the derivation of a master equation beyond such situation has been more elusive. 

In the projection operator techniques, a projection superoperator ${\mathcal P}$ is defined such that ${\mathcal P}\rho$ captures the relevant part of the total density matrix. Then an approximated master equation is obtained for ${\mathcal P}\rho$ based on a perturbative expansion with respect to the coupling between the (relevant part of the) system and the environment. The standard choice is to consider a projection superoperator such that ${\mathcal P}\rho=\rho_S(t)\otimes\rho_B$, where $\rho_S(t)=\Ttr_B{\rho}(t)$ \cite{breuer2004}, and then follow the Nakajima-Zwanig method \cite{nakajima1958,zwanzig1960}, that leads to an equation for ${\mathcal P}\rho$ which contains a time integration over the past history of the system. Alternatively, one may follow the so called time-convolutionless projection operator technique (TCL) which gives rise to an equation that is local in time \cite{breuer2006b,breuerbook}. 
Another possibility is to consider that the relevant part of the dynamics can be described with a correlated system-environment state, rather than a tensor product state $\rho_S(t)\otimes\rho_B$. In this regard, there is a class of projection superoperators that project the system into correlated system-environment states \cite{breuer2006b}, so that then TCL technique can be applied to calculate the evolution of ${\mathcal P}\rho$, where ${\mathcal P}$ is this class of projection operators. This approach was considered in \cite{esposito2003} and \cite{budini2005} by using two different types of correlated projector operators to derive two different types of master equations. 
An inconvenience of the projection operator techniques is that they rely strongly on the specific choice of a suitable set of projection operators, an election that might depend on the particular situation and be difficult to make for some cases.

In this paper we propose a method to derive a master equation without the use of any approximation, and considering an arbitrary system Hamiltonian $H_S$, which can in principle contain anharmonicities giving rise to a non-linear spectrum. We base our derivation on the so-called Stochastic Liouville von-Neumann (SLN) equation \cite{stockburger2002,wallsbook}, which describes exactly the dynamics of an oscillator coupled to an environment in thermal equilibrium, according to the Hamiltonian
\begin{eqnarray}
H_{SB}&=&H_S+\sum_{\lambda} \hbar \omega_\lambda a^{\dagger}_\lambda a_\lambda+\sum_{\lambda}\hbar \hat{g}_\lambda (a_\lambda +a^\dagger_\lambda)X,
\label{HSLN}
\end{eqnarray}
where $X$ is the system hermitian coupling operator, $a_\lambda (a_\lambda^\dagger)$ are the annihilation (creation) operators of the oscillator $\lambda$ of the environment, and $\hat{g}_\lambda$ and $\omega_\lambda$ are respectively its coupling and frequency. 
The stochastic sampling in the SLN method has often convergence problems that have to be tacked in rather sophisticated ways \cite{koch2008,koch2010}. Indeed, the convergence of the SLN methods fails particularly at long times and for super-ohmic reservoirs. It would be very desirable to have a master equation corresponding to the SLN equation, and hence avoid or at least have some insight into such convergence problems that appear in some cases.

This paper contains two main results: First, it presents to our knowledge the first analysis that allows deriving a master equation without any approximation or assumption. Following this analysis, a master equation is derived that comes in terms of a infinite sum of terms that form a hierarchical series. Second, it provides a criterion to numerically determine if a particular problem is tractable by considering the reduced dynamics in the OQS Hilbert space. In detail, if the subsequent types of terms in the series have more and more weight, it can be concluded that the system dynamics cannot be described by considering only an equation in its reduced Hilbert space. In this case, other approaches involving the total system Hilbert space should be considered instead. If on the contrary, the subsequent types of terms of the series have less and less weight, one can safely truncate the equation and consider it to give a non-approximated result up to an error derived from the truncation. 

\section{The SLN equation}
According to the seminal paper \cite{stockburger2002}, a path integral can be expressed as a stochastic average, ${\mathcal M}_{\xi,\nu}$ of two propagators $K_{\xi,\nu}(\dots)$ and $K^*_{\xi,\nu}(\dots)$ that depend on the Gaussian noises $\xi$ and $\nu$, 
\bea
\rho(x_{f},x'_{f})&=&\int dx_i dx'_i {\mathcal M}_{\xi,\nu}[K_{\xi,\nu}(x_{f},x_i)K^*_{\xi,\nu}(x'_{f},x'_i)]\nonumber\\
&\times&\rho(x_i,x'_i,0).
\eea
The two Gaussian noises must fulfil the following statistical properties,
\bea
{\mathcal M}_{\xi,\nu}[\xi(t)\xi(t')]&=&\alpha_R(t-t');\nonumber\\
{\mathcal M}_{\xi,\nu}[\xi(t)\nu(t')]&=&-i\alpha_I(t-t')\theta(t-t');\nonumber\\
{\mathcal M}_{\xi,\nu}[\nu(t)\nu(t')]&=&0, 
\label{prop}
\eea
where ${\mathcal M}_{\xi,\nu}[\cdots]$ is a Gaussian average over the two different noises, and $\theta(t)$ is the Heaviside step function. 
In the above definitions, $\alpha_R(t)$ and $\alpha_I(t)$ correspond respectively to the real and imaginary parts of the correlation function corresponding to the Hamiltonian (\ref{HSLN}), 
\begin{eqnarray}
\alpha_T (t-\tau)&=&\sum_\lambda \hat{g}^2_\lambda \bigg[\coth\left(
{\frac{\omega_\lambda \beta}{2}}\right)\cos\left(\omega_\lambda (t-\tau)\right)\cr
&-&i\sin\left(\omega_\lambda (t-\tau)\right)\bigg],
\label{correl3chap3a}
\end{eqnarray}
where $\hat{g}_\lambda$ are the coupling constants of the original Hamiltonian, and $\beta=1/K_BT$, whith $k_B$ the Boltzmann constant and $T$ the environment temperature.

According to \cite{stockburger2002}, a similar procedure to the one in \cite{feynman1963} leads to a stochastic differential equation for the reduced density operator of a single trajectory (dependent on the noises $\xi(t)$ and $\nu(t)$), the so-called stochastic Liouville von Neumann (SLN) equation,
\bea
\frac{dP_{\xi,\nu}}{dt}&=&-\frac{i}{\hbar}[H_s,P_{\xi,\nu}]+\frac{i}{\hbar}\xi(t)[\hat{x},P_{\xi,\nu}]+\frac{i}{2}\nu(t)\{\hat{x},P_{\xi,\nu}\}\cr
\label{LVN}
\eea 
Here we have neglected a re-normalization factor which is not relevant for the present derivation. 
This equation, valid for environments in thermal equilibrium, allows to compute different noise trajectories for $P_{\xi,\nu}$, such that the reduced density operator can be obtained as 
\bea
\rho_s(t)={\mathcal M}_{\xi,\nu}[P_{\xi,\nu}].
\label{average}
\eea
The above equation (\ref{LVN}) can be re-written as two stochastic equations for two different stochastic wave vectors $|\psi_t^1\rangle$ and $|\psi_t^2\rangle$, 
\bea
&&\frac{d|\psi^1_t\rangle}{dt}=-iH_S|\psi^1_t\rangle+ i\xi(t)L|\psi^1_t\rangle+i\frac{1}{2}\nu(t)L|\psi^1_t\rangle,\cr
&&\frac{d|\psi^2_t\rangle}{dt}=-iH_S|\psi^2_t\rangle+ i\xi^*(t)L|\psi^2_t\rangle-i\frac{1}{2}\nu(t)^*L|\psi^2_t\rangle.
\label{SSEexact3}
\eea
such that $P_{\xi,\nu}=|\psi^1_t\rangle\langle\psi^2_t|$ \cite{koch2010}. In the above equations, we have have ignored the re-normalization factor appearing in \cite{koch2010}, since it is not relevant for the present derivation. 


\section{Rexpressing the noises in (\ref{LVN})}

To solve equations (\ref{LVN}) or (\ref{SSEexact3}), the noises $\xi(t)$ and $\nu(t)$ are numerically built in terms of two different response functions \cite{koch2010}. We propose here to follow a different approach by defining these noises in terms of Gaussian white noises, coupling operators $g_\lambda$, and environment frequencies $\omega_\lambda$ of the Hamiltonian (\ref{HSLN}). 
The newly defined noises have the following form, 
\bea
\xi(t)&=&z_{0}(t)+z^*_{0}(t)+2z_{1}(t);\cr
\nu(t)&=&iz^*_{1}(t)
\label{newnoise}
\eea
%
Where we have defined, 
\bea
z_{0}(t)&=&\frac{1}{\sqrt{2}}\sum_\lambda g_\lambda z_{0\lambda} e^{-i\omega_\lambda t},\cr
z^*_{0}(t)&=&\frac{1}{\sqrt{2}}\sum_\lambda g_\lambda z^*_{0\lambda} e^{i\omega_\lambda t},\cr
z_{1}(t)&=&\sum_\lambda f_\lambda z_{1\lambda}e^{-i\omega_\lambda t}\cr,
z^*_{1}(t)&=&\sum_\lambda h^*_\lambda z^*_{1\lambda} e^{i\omega_\lambda t}.
\label{newnoise2}
\eea
with $z_{j\lambda}$, for $j=0,1$ being two different (i.e. uncorrelated) complex quantities with properties ${\mathcal M}[z_{j\lambda}]=0$, ${\mathcal M}[z_{j\lambda}, z_{i\lambda'}]=0$, ${\mathcal M}[z_{j\lambda}z^*_{i\lambda'}]=\delta_{ji}\delta_{\lambda,\lambda'}$. Note that while $z^*_{0}(t)=(z_0(t))^*$, $z^*_1(t)\neq(z_1(t))^*$. Importantly, the above defined noises obey the same statistical properties as (\ref{prop})
\bea
&&{\mathcal M}_{\xi,\nu}[\xi(t)\xi(t')]=\sum_\lambda g_\lambda^2\cos(\omega_\lambda(t-\tau))=\alpha_R(t-t');\nonumber\\
&&{\mathcal M}_{\xi,\nu}[\xi(t)\nu(t')]=i2\sum_\lambda f_\lambda h^*_\lambda e^{-i\omega_\lambda (t-t')}=
i{\mathcal X}_R(t-t');\nonumber\\
&&{\mathcal M}_{\xi,\nu}[\nu(t)\nu(t')]=0.
\label{prop3}
\eea
where ${\mathcal X}_R(t-t')=2\int^\infty_{-\infty}d\omega f(\omega)h^*(\omega)e^{-i\omega t}=-2\alpha_I(t-t')\theta(t-t')$. The advantage of expressing the noises as (\ref{newnoise}), is that the Gaussian average can now be explicitly written as ${\mathcal M}_{\xi,\nu}[\cdots]={\mathcal M}[\cdots]=\int d^2z_0 e^{-|z_0|^2}\int d^2z_1 e^{-|z_1|^2}\cdots$, where $z_j=\{z_{j0}, z_{j1}, \cdots z_{j\lambda},\cdots\}$ denotes set of complex numbers corresponding to each of the $\lambda$ harmonic oscillators. This will allow us, in the following sections, to carry on analytically instead of numerically, the average ${\mathcal M}_{\xi,\nu}[\cdots]$ that permits to obtain the reduced density matrix (\ref{average}) or its evolution equation from (\ref{LVN}).

The functions $f(\omega)$ and $h(\omega)$ in (\ref{newnoise2}) can be chosen as $f(\omega)=\sqrt{-\frac{1}{2}{\mathcal X}_R(\omega)}$, and $h(\omega)=\bigg(\frac{-{\mathcal X}_R(\omega)}{2f(\omega)}\bigg)^*$ for instance \cite{koch2010}. However, their value is unimportant for the present derivation for which the only relevant quantity is the correlation function (\ref{correl3chap3a}). 
Note also that we should define $g_\lambda=\frac{1}{\sqrt{2}}\hat{g}_\lambda (\sqrt{1+2N(\omega_\lambda)})$. In that way, the real part of the correlation function can be written in terms of $g_\lambda^2=\frac{1}{2}\hat{g}^2_\lambda\coth\left(
{\frac{\omega_\lambda \beta}{2}}\right)=\frac{1}{2}\hat{g}^2_\lambda(2N(\omega_\lambda)+1)$, with $N(\omega)=[\exp(\omega\beta)-1]^{-1}$ the average thermal number of quanta in the mode $\omega$. The quantity $\beta=1/(K_BT)$, with $T$ is the temperature and $K_B$ the Boltzmann constant. 

%
%

\section{Deriving a master equation}

In order to derive a master equation, we need to compute $\frac{d{\mathcal M}[P]}{dt}$, where for simplicity in the notation we have renamed $P=P_{\xi,\nu}$. To this order, we take into account that $\frac{d{\mathcal M}[P]}{dt}={\mathcal M}\large[\frac{dP}{dt}\large]$, and consider the SLN equation (\ref{LVN}) for $P$. 
In order to have a more compact notation, we re-express the equation (\ref{LVN}) in the interaction picture with respect to the system and within a Liouvillian form,
\bea
\frac{dP^v_t}{dt}&=&{\mathcal L}(t)P^v_t
\eea
where ${\mathcal L}(t)$ is a super-operators acting over the the stochastic projector $P_t$ flattened as a vector, denoted as $P^v_t$. In detail, ${\mathcal L}(t)P^v_t=\frac{i}{\hbar}\xi(t)[X(t),P_{\xi,\nu}]+\frac{i}{2}\nu(t)\{X(t),P_{\xi,\nu}\}$, where $X(t)=e^{iH_St}\hat{x}e^{-iH_St}$. The equation above can be re-written as 
\bea
\frac{dP^v_t}{dt}&=&\sum_{j=0,1}\sum_\lambda L^{0}_{j\lambda} (t)z_{j\lambda} P^v_t\cr
&+&\sum_\lambda L^{*}_{j\lambda}(t)z^*_{j\lambda} P_t
\label{project2}
\eea
where $L^{0}_{j\lambda} (t)=\frac{\partial {\mathcal L}(t)}{\partial z_{j\lambda}}$, and $L^{*}_{j\lambda} (t)=\frac{\partial {\mathcal L}(t)}{\partial z^*_{j\lambda}}$ are new super-operators. Note that because of the interaction image, these quantities are rotated with respect to $H_S$, and do no longer depend on any noise $z_{i\lambda}$ or $z^*_{i\lambda}$. 
The analytical average of the former equation, leads to the evolution of the reduced density matrix of the system as 
\bea
\frac{d\rho_s^v(t)}{dt}={\mathcal M}\large[\frac{dP^v_t}{dt}\large]&=&\sum_{j=0,1}\sum_\lambda L^{0}_{j\lambda} (t){\mathcal M}\large[z_{j\lambda} P^v_t\large]\cr
&+&\sum_\lambda L^{*}_{j\lambda}(t){\mathcal M}\large[z^*_{j\lambda} P_t\large].
\label{project21}
\eea
Unfortunately this is an open equation, in the sense that the derivative of $\rho_s(t)$ does not depend only on $\rho_s$ as expected for a proper master equation, but it also depends on the unknown quantities ${\mathcal M}[z_{j\lambda} P^v_t]$ and ${\mathcal M}[z^*_{j\lambda} P^v_t]$. Considering that ${\mathcal M}[\cdots]$ are integrals over a Gaussian measure, these quantities can be re-written as 
\bea
&&{\mathcal M}[z_{j\lambda} P^v_t]={\mathcal M}\bigg[\frac{\partial P^v_t}{\partial z^*_{j\lambda}}\bigg]=A^*_\lambda\cr
&&{\mathcal M}[z^*_{j\lambda} P^v_t]={\mathcal M}\bigg[\frac{\partial P^v_t}{\partial z_{j\lambda}}\bigg]=A^0_\lambda
\eea
where the right hand side of the expressions has been obtained by considering an integration by parts of the left hand side. 
As one can see, the evolution of $P(t)$ is intimately related to the evolution of $\frac{\partial P^v_t}{\partial z^*_{j\lambda}}$ and $\frac{\partial P^v_t}{\partial z_{j\lambda}}$. The evolution equation for $\rho_s^v$ is then written in the form 
\bea
\frac{d\rho^v_s(t)}{dt}=\sum_{j=0,1}\sum_\lambda L^{0}_{j\lambda} (t) A^*_\lambda
+\sum_\lambda L^{*}_{j\lambda}(t)A^0_\lambda.
\label{masteropen}
\eea
This is still an open equation that depends on two unknown quantities $A^*_\lambda(t)$ and $A^0_\lambda(t)$. To proceed further, we need to re-express these quantities in terms of $\rho^v_s(t)$. The first step is to calculate their evolution equation, taking into account the consistency conditions $\frac{d}{dt}\frac{\partial P}{\partial z_{j\lambda}}=\frac{\partial}{\partial z_{j\lambda}}\frac{dP}{dt}$, and $\frac{d}{dt}\frac{\partial P}{\partial z^*_{j\lambda}}=\frac{\partial}{\partial z^*_{j\lambda}}\frac{dP}{dt}$, together with equation (\ref{project2}),
\bea
\frac{\partial}{\partial z_{j\lambda}}\frac{dP^v_t}{dt}&=&L^{0}_{j\lambda} (t)P^v_t+\sum_{j'=0,1}\sum_{\lambda'} L^{0}_{j'\lambda'} (t)z_{j'\lambda'} \frac{\partial P^v_t}{\partial z_{j\lambda}}\cr
&+&\sum_{\lambda'} L^{*}_{j'\lambda'}(t)z^*_{j'\lambda'} \frac{\partial P^v_t}{\partial z_{j\lambda}}\cr
\frac{\partial}{\partial z^*_{j\lambda}}\frac{dP^v_t}{dt}&=&L^{*}_{j\lambda} (t)P^v_t+\sum_{j=0,1}\sum_{\lambda'} L^{0}_{j\lambda'} (t)z_{j'\lambda'} \frac{\partial P^v_t}{\partial z^*_{j\lambda}}\cr
&+&\sum_{\lambda'} L^{*}_{j'\lambda'}(t)z^*_{j'\lambda'} \frac{\partial P^v_t}{\partial z^*_{j\lambda}}
\label{project3}
\eea
If we now compute the average ${\mathcal M}[\cdots]$ of the former equation, we will find 
\bea
\frac{dA_{\lambda}^0}{dt}&=&L_{j\lambda}^0 (t)\rho^v_s(t)+\sum_{j'=0,1}\sum_{\lambda'} L^{0}_{j'\lambda'} (t)B_{\lambda'\lambda}^{*0}\cr
&+&\sum_{j'=0,1}\sum_{\lambda'} L^{*}_{j'\lambda'}(t)B_{\lambda'\lambda}^{00}\cr
\frac{dA_{\lambda}^*}{dt}&=&L_{j\lambda}^* (t)\rho^v_s(t)+\sum_{j'=0,1}\sum_{\lambda'} L^{0}_{j'\lambda'} (t)B_{\lambda'\lambda}^{**}\cr
&+&\sum_{j'=0,1}\sum_{\lambda'} L^{*}_{j'\lambda'}(t)B_{\lambda'\lambda}^{0*}.
\label{project5}
\eea
which depends on terms of the form 
\bea
&&{\mathcal M}[z_{j'\lambda'} \frac{\partial P^v_t}{\partial z_{j\lambda}}]={\mathcal M}\bigg[\frac{\partial}{\partial z^*_{j'\lambda'}}\frac{\partial P^v_t}{\partial z_{j\lambda}}\bigg]=B^{*0}_{\lambda'\lambda}\cr
&&{\mathcal M}[z_{j'\lambda'} \frac{\partial P^v_t}{\partial z^*_{j\lambda}}]={\mathcal M}\bigg[\frac{\partial}{\partial z^*_{j'\lambda'}}\frac{\partial P^v_t}{\partial z^*_{j\lambda}}\bigg]=B^{**}_{\lambda'\lambda}\cr
&&{\mathcal M}[z^*_{j'\lambda'} \frac{\partial P^v_t}{\partial z_{j\lambda}}]={\mathcal M}\bigg[\frac{\partial}{\partial z_{j'\lambda'}}\frac{\partial P^v_t}{\partial z_{j\lambda}}\bigg]=B^{00}_{\lambda'\lambda}\cr
&&{\mathcal M}[z^*_{j'\lambda'} \frac{\partial P^v_t}{\partial z^*_{j\lambda}}]={\mathcal M}\bigg[\frac{\partial}{\partial z_{j'\lambda'}}\frac{\partial P^v_t}{\partial z^*_{j\lambda}}\bigg]=B^{0*}_{\lambda'\lambda}.
\eea
For the shake of simplicity, we note that in the quantities $A_\lambda^\beta$ and $B_{\lambda\lambda'}^{\beta\beta'}$ ($\beta,\beta'=*,0$), we skip the indexes $j$ and $j'j$ respectively.
To reach even a higher order, we need to compute the evolution of the four quantities 
$\frac{\partial}{\partial z^*_{j\lambda'}}\frac{\partial P^v_t}{\partial z_{j\lambda}}$, $\frac{\partial}{\partial z^*_{j\lambda'}}\frac{\partial P^v_t}{\partial z^*_{j\lambda}}$, $\frac{\partial}{\partial z_{j\lambda'}}\frac{\partial P^v_t}{\partial z_{j\lambda}}$, and $\frac{\partial}{\partial z_{j\lambda'}}\frac{\partial P^v_t}{\partial z^*_{j\lambda}}$. To this order, we consider the above introduced consistency condition, and the evolution equations (\ref{project3}) to get the following,
\bea
&&\frac{\partial^2}{\partial z_{j'\lambda'}\partial z_{j\lambda}}\frac{dP}{dt}=L^{0}_{j\lambda} (t)\frac{\partial P^v_t}{\partial z_{j'\lambda'}}+L^{0}_{j'\lambda'} (t)\frac{\partial P^v_t}{\partial z_{j\lambda}}\cr
&+&\sum_{j'=0,1}\sum_{\lambda''} \bigg(L^{0}_{j'\lambda''} (t)z_{j'\lambda''} +L^{*}_{j'\lambda''}(t)z^*_{j'\lambda'}\bigg)\frac{\partial^2P^v_t}{\partial z_{j'\lambda'}\partial z_{j\lambda}}\cr
&&\frac{\partial^2}{\partial z^*_{j'\lambda'}\partial z_{j\lambda}}\frac{dP}{dt}=L^{0}_{j\lambda} (t)\frac{\partial P^v_t}{\partial z^*_{j'\lambda'}}+L^{*}_{j'\lambda'} (t)\frac{\partial P^v_t}{\partial z_{j\lambda}}\cr
&+&\sum_{j'=0,1}\sum_{\lambda''} \bigg(L^{0}_{j'\lambda''} (t)z_{j'\lambda''} +L^{*}_{j'\lambda''}(t)z^*_{j'\lambda''}\bigg)\frac{\partial^2P^v_t}{\partial z^*_{j'\lambda'}\partial z_{j\lambda}}\cr
&&\frac{\partial^2}{\partial z_{j'\lambda'}\partial z^*_{j\lambda}}\frac{dP^v_t}{dt}=L^{*}_{j\lambda} (t)\frac{\partial P^v_t}{\partial z_{j'\lambda'}}+L^{0}_{j'\lambda'} (t)\frac{\partial P^v_t}{\partial z^*_{j\lambda}}\cr
&+&\sum_{j'=0,1}\sum_{\lambda''} \bigg(L^{0}_{j'\lambda''} (t)z_{j'\lambda''} +L^{*}_{j'\lambda''}(t)z^*_{j'\lambda''}\bigg)\frac{\partial^2 P^v_t}{\partial z_{j'\lambda'}\partial z^*_{j\lambda}}\cr
&&\frac{\partial^2}{\partial z^*_{j'\lambda'}\partial z^*_{j\lambda}}\frac{dP^v_t}{dt}=L^{*}_{j\lambda} (t)\frac{\partial P^v_t}{\partial z^*_{j'\lambda'}}+L^{*}_{j'\lambda'} (t)\frac{\partial P^v_t}{\partial z^*_{j\lambda}}\cr
&+&\sum_{j'=0,1}\sum_{\lambda''} \bigg(L^{0}_{j'\lambda''} (t)z_{j'\lambda''} +L^{*}_{j'\lambda''}(t)z^*_{j'\lambda''}\bigg)\frac{\partial^2P^v_t}{\partial z^*_{j'\lambda'}\partial z^*_{j\lambda}}\cr
\label{project6}
\eea
Computing the stochastic average of the former equations gives rise to the following set of equations,
\bea
\frac{dB_{\lambda'\lambda}^{00}}{dt}&=&L^{0}_{j\lambda} (t)A_{j'\lambda'}^0+L^{*}_{j'\lambda'} (t)A_{\lambda}^0\cr
&+&\sum_{j'=0,1}\sum_{\lambda'} \bigg(L^{0}_{j'\lambda'} (t)C^{*00}_{\lambda''\lambda'\lambda} +L^{*}_{j'\lambda'}(t)C^{000}_{\lambda''\lambda'\lambda}\bigg)\cr
\frac{dB_{\lambda'\lambda}^{*0}}{dt}&=&L^{0}_{j\lambda} (t)A_{j'\lambda'}^*+L^{0}_{j'\lambda'} (t)A_{\lambda}^0\cr
&+&\sum_{j'=0,1}\sum_{\lambda'} \bigg(L^{0}_{j'\lambda'} (t)C^{**0}_{\lambda''\lambda'\lambda}+L^{*}_{j'\lambda'}(t)C^{0*0}_{\lambda''\lambda'\lambda}\bigg)\cr
\frac{dB_{\lambda'\lambda}^{**}}{dt}&=&L^{*}_{j\lambda} (t)A_{j'\lambda'}^*+L^{*}_{j'\lambda'} (t)A_{\lambda}^*\cr
&+&\sum_{j'=0,1}\sum_{\lambda'} \bigg(L^{0}_{j'\lambda'} (t)C^{***}_{\lambda''\lambda'\lambda}+L^{*}_{j'\lambda'}(t)C^{0**}_{\lambda''\lambda'\lambda}\bigg)\cr
\frac{dB_{\lambda'\lambda}^{0*}}{dt}&=&L^{*}_{j\lambda} (t)A_{j'\lambda'}^0+L^{0}_{j'\lambda'} (t)A_{\lambda}^*\cr
&+&\sum_{j'=0,1}\sum_{\lambda'} \bigg(L^{0}_{j'\lambda'} (t)C^{*0*}_{\lambda''\lambda'\lambda}+L^{*}_{j'\lambda'}(t)C^{00*}_{\lambda''\lambda'\lambda}\bigg)\cr
\label{project7}
\eea
where the $C^{\alpha\beta\gamma}_{\lambda''\lambda'\lambda}$, with $\alpha,\beta,\gamma=0,*$ are given by partial derivatives of the projector over $z^\alpha_{j''\lambda''}$, $z^\beta_{j'\lambda'}$ and $z^\gamma_{j\lambda}$. 

The equations (\ref{masteropen}), (\ref{project5}) and (\ref{project7}) form an open system of mutually dependent equations. Because of this structure, the evolution of $\rho^v_S(t)$ can be written in terms of a complex series of terms that only depend on $\rho^v_S$ at past times. The first two terms of the series can be obtained as follows:
\begin{itemize}
\item (i) First integrate analytically (\ref{project5}), obtaining a formal solution for $A_\lambda^0(t)$, and $A_\lambda^*(t)$, which comes in terms of different $B$s,
\bea
A_{\lambda}^0(t)&=&A_{\lambda}^0(0)+\int_0^t ds \bigg[L_{j\lambda}^0 (s)\rho^v_s(s)+\sum_{j'=0,1}\sum_{\lambda'}\bigg(\cr &\times&L^{0}_{j'\lambda'} (s)B_{\lambda'\lambda}^{*0}(s)+L^{*}_{j'\lambda'}(t)B_{\lambda'\lambda}^{00}\bigg)\bigg]\cr
A_{\lambda}^*(t)&=&A_{\lambda}^*(0)+\int_0^t ds \bigg[L_{j\lambda}^* (s)\rho^v_s(s)+\sum_{j'=0,1}\sum_{\lambda'}\bigg(\cr &\times& L^{0}_{j'\lambda'} (s)B_{\lambda'\lambda}^{**}(s)+ L^{*}_{j'\lambda'}(s)B_{\lambda'\lambda}^{0*}(s)\bigg)\bigg].
\label{intproject5}
\eea
Note that all initial states of $A$, $B$, $C$, etc are zero; 

\item (ii) In a similar way, compute the different $B$s by analytically integrating (\ref{project7}). From this, an equation (21-b) is obtained, that expresses the $B$s in terms of $A$s and $C$s; 

\item (iii) Insert the values of $A$s given by (\ref{intproject5}) in this equation giving rise to an equation here labeled as (21-c). 

\item (iv) Since (21-c) will come again in terms of $B$s, replace on its right hand side the expression of $B$ given by (21-b). 
\end{itemize} 
At the end, an equation for the different $B$s is obtained which comes only in terms of $\rho_s^v$ at different times, and higher order terms $C^{\alpha\beta\gamma}_{\lambda''\lambda'\lambda}$. Neglecting these higher order terms, this equation for $B$s shall be inserted in (\ref{intproject5}). The resulting expression for $A$ shall be inserted in the r.h.s. of (\ref{masteropen}), giving the following master equation
\bea
&&\frac{d\rho_s^v(t)}{dt}=\sum_{j=0,1}\int_0^t ds \bigg\{\hat{\alpha}^{0*}_{jj}(t-s) {\mathcal L}_j^0(t) {\mathcal L}_j^*(s)\cr
&+&\int_0^t ds \hat{\alpha}^{*0}_{jj}(t-s) {\mathcal L}_j^*(t) {\mathcal L}_j^0(s)\bigg\}\rho^v(s)\cr
&+&
\sum_{\substack{\alpha,\beta,\gamma,\eta\\
\alpha\neq\beta,
\gamma\neq\eta}}
\int_0^t ds \int_0^s ds'\int_0^{s'} ds''\hat{\alpha}^{\alpha\beta}_{jj}(t-s')\alpha^{\gamma\eta}_{jj}(s-s'')\cr
&\times&{\mathcal U}^{\alpha,\beta,\gamma,\eta}_{jj'jj'}(t,s,s',s'')\rho_s^v(s)\cr
&+&\sum_{\substack{\alpha,\beta,\gamma,\eta\\
\alpha\neq\beta,
\gamma\neq\eta}}
\int_0^t ds \int_0^s ds' \int_0^{s'} ds''\alpha^{\alpha\beta}_{jj}(t-s'')\hat{\alpha}^{\gamma\eta}_{jj}(s-s')\cr
&\times&{\mathcal U}^{\alpha,\gamma,\eta,\beta}_{jj'j'j}(t,s,s',s'')\rho_s^v(s)+\cdots
\label{masteropen2}
\eea
where $\alpha,\beta,\eta,\nu=0,*$, and $j,i,n,m=0,1$, and $\rho^v(t)$ is the reduced density matrix of the system flattened as a vector. Note that because of the restrictions of the sums, we are left with terms proportional to $\alpha^{0*}_{jj}\alpha^{0*}_{jj}$, $\alpha^{0*}_{jj}\alpha^{*0}_{jj}$, $\alpha^{*0}_{jj}\alpha^{*0}_{jj}$, or $\alpha^{*0}_{jj}\alpha^{0*}_{jj}$. We have also defined ${\mathcal U}^{\alpha,\beta,\gamma,\eta}_{jinm}(t,s,s',s'')={\mathcal L}_j^\alpha(t){\mathcal L}_i^\beta(s){\mathcal L}_n^\gamma(s'){\mathcal L}_m^\eta(s'')$. The super-operators appearing in this expression are
${\mathcal L}_0^0(t)={\mathcal L}_0^*(t)={\mathcal L}_M(t)$, ${\mathcal L}_1^0(t)=2{\mathcal L}_M(t)$, and ${\mathcal L}_1^*(t)={\mathcal L}^c_M(t)$, where 
\bea
&&{\mathcal L}_M(t)A^v=i[X(t),A(t)];\cr
&&{\mathcal L}^c_M(t)A^v=-\frac{1}{2}\{X(t),P(t)\},
\label{values}
\eea
As before, the quantity $A^v$ in the l.h.s. is a the vector with size $2^{2N}$ flattened from a matrix $A$ with size $2^N\times 2^N$, where $N$ is the OQS dimension. Finally, in this expression we have also settled
\bea
\hat{\alpha}^{0*}_{00}(t,s)&=&\frac{1}{2\hbar}\sum_\lambda g_\lambda^2 e^{-i\omega_\lambda(t-s)}\cr
\hat{\alpha}^{0*}_{11}(t,s)&=&\frac{i}{2\hbar}\sum_\lambda f_\lambda h_\lambda^* e^{-i\omega_\lambda(t-s)}\cr
&=&-2\alpha_I(t-s)\theta(t-s).
\label{correlationsM}
\eea
Note also that 
\bea
\hat{\alpha}^{*0}_{00}(t,s)&=&\frac{1}{2\hbar}\sum_\lambda g_\lambda^2 e^{i\omega_\lambda(t-s)}\cr
\hat{\alpha}^{*0}_{11}(t,s)&=&\frac{i}{2\hbar}\sum_\lambda f_\lambda h_\lambda^* e^{-i\omega_\lambda(-(t-s))}\cr
&=&-2\alpha_I(s-t)\theta(s-t).
\eea

\begin{figure}[ht]
\centerline{\includegraphics[width=0.45\textwidth]{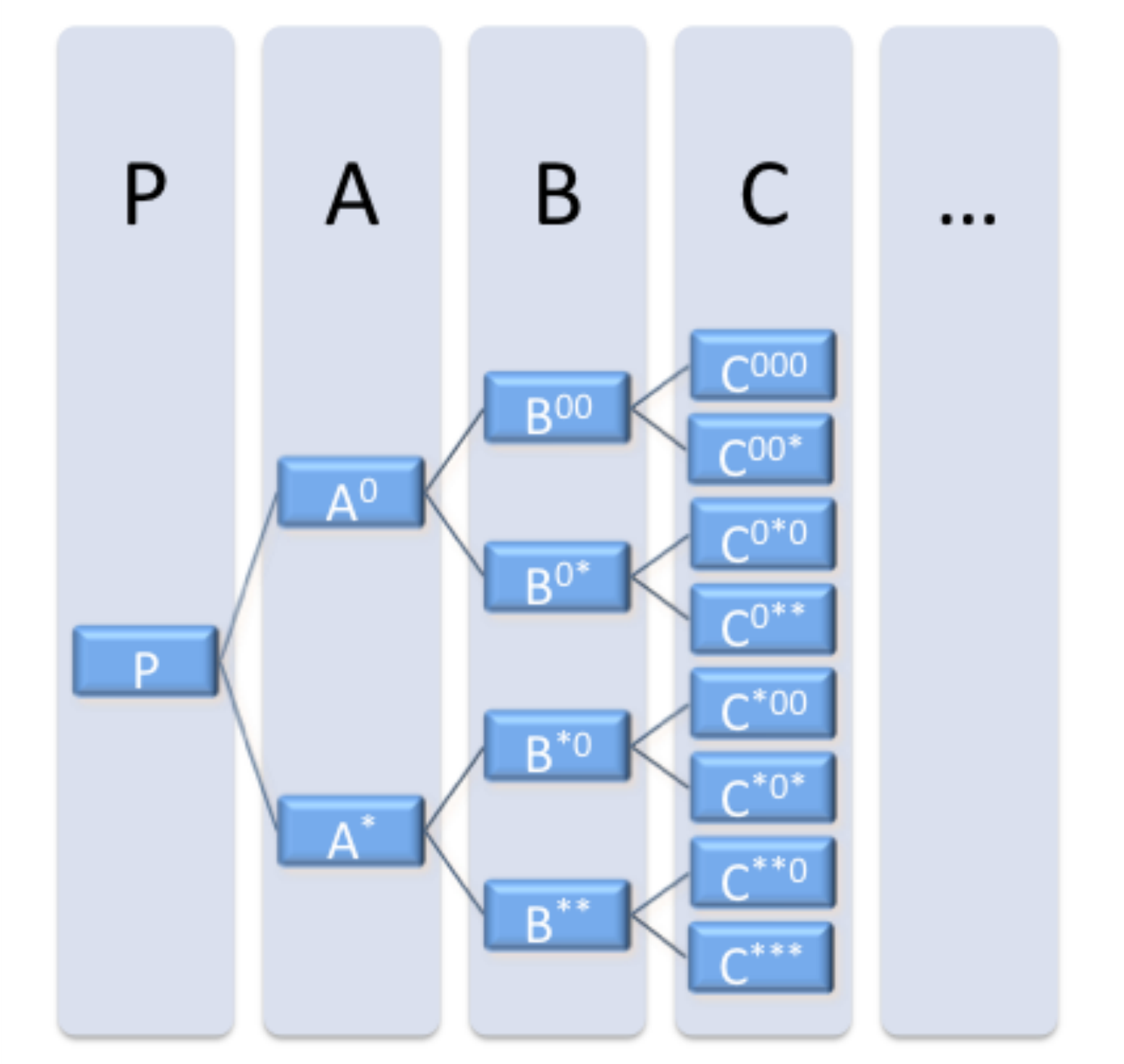}}
\caption{Different levels of the hierarchy, corresponding to the average ${\mathcal M}[\cdots]$ of different orders of the derivative of the stochastic projector $P_t$ with respect to the quantities $z_0$, $z^*_0$, $z_1$, and $z^*_1$. The zero order derivative is given by the equation (\ref{masteropen}), while the first and the second order are given by (\ref{project5}) and (\ref{project7}) respectively. \label{esquema}}
\end{figure}

The number of members of each class, $A^{\alpha}_{\lambda}$ (referred here as first order), $B^{\alpha\beta}_{\lambda'\lambda}$ (second order), $C^{\alpha\beta\gamma}_{\lambda''\lambda'\lambda}$ (third order), etc, are $P\times 2^k$, where $k$ is the order number, and $P$ is the number of complex noises, in our case $P=2$ (see Fig. (\ref{esquema})). 
Thus, the master equation (\ref{masteropen2}) is written in terms of an infinite series of terms, wherein the first two terms are in the form of an integral over $\rho_s$ involving one correlation function (\ref{correl3chap3a}), and the following eight terms are in the form of a triple integral over $\rho_s$ involving two correlation functions. The next terms will come in the form of five integrals involving three correlation functions, and so on and so forth. 

It is important to take into account the convergence properties of the series. If the system parameters are such that the relative weight of the terms involving a single integral is smaller than that of the terms involving a triple integral, then the case is not tractable because the equation cannot be truncated. It is worth to notice at this point that the above-described hierarchical structure does not appear because of the use of any expansion approximation, such as the weak coupling. Rather, it is an structure \textit{intrinsically attached to the OQS problem}. Interestingly, a similar hierarchical structure was obtained in \cite{alonso2007} within the Heisenberg approach. Hence, for systems or parameter ranges where the hierarchy structure does not converge (i.e. successive terms are not smaller than the previous ones), one could conclude that these particular instances can not be correctly dealt with master equations. Indeed, in these situations the dynamics of the open quantum system appear to alter the environment so dramatically, that obtaining a closed equation within the system's Hilbert space is not feasible in practice. In contrast, for these problems where the subsequent terms are less important, one can safely truncate the equation. In such cases, the truncated equation gives the dynamics of the reduced density matrix \textit{without any approximation}, only with an error that can be estimated by the magnitude of the terms that have been discarded. 

\section{Weak coupling and Markov limit}
For consistency, let us check that equation (\ref{masteropen2}) coincides with the master equation up to second order in the coupling parameter $g$ derived within other approaches. Indeed, each correlation function $\alpha_{jj}^{\alpha\beta}$ will be of the order $g^2$, so that if we just keep second order terms, the equation takes the form 
\bea
&&\frac{d\rho_s(t)}{dt}=\sum_{j=0,1}\int_0^t ds \bigg\{\hat{\alpha}^{0*}_{jj}(t-s) {\mathcal L}_j^0(t) {\mathcal L}_j^*(s)\cr
&+&\int_0^t ds \hat{\alpha}^{*0}_{jj}(t-s) {\mathcal L}_j^*(t) {\mathcal L}_j^0(s)\bigg\}\rho_s(s).
\eea
Replacing the values ${\mathcal L}_0^0(t)={\mathcal L}_0^*(t)={\mathcal L}_M(t)$, ${\mathcal L}_1^0(t)=2{\mathcal L}_M(t)$, and ${\mathcal L}_1^*(t)={\mathcal L}^c_M(t)$, we find 
\bea
&&\frac{d\rho_s(t)}{dt}=-\int_0^t ds \bigg\{\Rre\bigg[\hat{\alpha}^{0*}_{00}(t-s)\bigg] {\mathcal L}_M(t) {\mathcal L}_M(s)\cr
&+&i\int_0^t ds \hat{\alpha}^{0*}_{11}(t-s) {\mathcal L}_M(t) {\mathcal L}^c_M(s)\cr
&+&i\int_0^t ds \hat{\alpha}^{*0}_{11}(t-s) {\mathcal L}^c_M(t) {\mathcal L}_M(s)\bigg\}\rho_s(s),
\eea
with ${\mathcal L}_M(t)$ and ${\mathcal L}^c_M(t)$ defined in (\ref{values}).
Re-written the resulting equation in its matrix form, it is identical to the time-convoluted master equation \cite{gaspard1999b,breuerbook} up to second order in $g$,
\begin{eqnarray}
&&\frac{d{\rho_s}(t)}{dt}=-
\int_0^td{\tau}\alpha_T (t-\tau)X(t)X(\tau)\rho_s(s) \nonumber\\
&-& \int_0^td{\tau}\alpha_T^*(t-\tau)\rho_s (s) X(\tau) X(t) \cr
&+&\int_0^td{\tau}\alpha_T(t-\tau)X(\tau)\rho_s(s)X(t) 
+ \int_0^td{\tau}\alpha_T^* (t-\tau)\cr
&\times&X(t)\rho_s(s)X(\tau)+{\mathcal O}(g^4), 
\label{masterconv}
\end{eqnarray}
As noted above, the equation is convoluted, because the reduced density matrix is within the integral, and not local in time as in the so called convolution-less master equations \cite{breuerbook}. However, taking into account that in interaction image with respect to the system $\rho_s(s)=\rho_s(t)+\mathcal{O}(g^2)$, and that the $\rho_s(s)$ appears already in second order terms, one can write the former equation as 
\begin{eqnarray}
&&\frac{d{\rho_s}(t)}{dt}=-
\int_0^td{\tau}\alpha_T (t-\tau)X(t)X(\tau)\rho_s(t) \nonumber\\
&-& \int_0^td{\tau}\alpha_T^*(t-\tau)\rho_s (t) X(\tau) X(t) \cr
&+&\int_0^td{\tau}\alpha_T(t-\tau)X(\tau)\rho_s(t)X(t) 
+ \int_0^td{\tau}\alpha_T^* (t-\tau)\cr
&\times&X(t)\rho_s(t)X(\tau)+{\mathcal O}(g^4), 
\label{masterconv2}
\end{eqnarray}
This is already a time local second order master equation, identical to the one derived following the TCL projection operator technique \cite{breuerbook}, and the second order SSEs \cite{yu1999,devega2005a}.

If one considers the Markov limit of the master equation (\ref{masteropen2}), by considering $\alpha(t)=\Gamma\delta(t)$, the terms involving three integrals (and two correlation functions) vanish, while the terms involving a single integral give rise to the well-known Lindblad equation \cite{lindblad1976}.
\begin{eqnarray}
\frac{d{\rho_s}(t)}{dt}&=&-\Gamma (X(t)X(t)\rho_s(t)+\rho_s (t) X(t) X(t) )\cr
&+&2X(t)\rho_s(t)X(t) \label{lind}
\end{eqnarray}
\section{Conclusions and perspectives}

We have derived an evolution equation for the reduced density operator of the system without considering any approximation. Such equation is written in terms of a hierarchy of terms that involve an increasing number of integrals containing the reduced density operator of the system. 

The derivation suggests that it may not be possible to deal with certain problems using a master equation approach, or considering the dynamics in the reduced Hilbert space of the system. These problems may therefore have to be treated by considering an evolution equation for the total system (i.e. OQS and environment). In this regard, the hierarchical structure of the resulting evolution equation leads to a specific criterion to determine numerically whether the master equation approach (namely calculating the evolution of the reduced density matrix of the system) is indeed suitable or not for a particular problem. This criteria is based in measuring the relative weight of the successive types of terms within equation (\ref{masteropen2}). If the relative weight of the second type of terms (i.e. the terms involving three integrals) is larger than that of the first type of terms (the terms involving a single integral), then it can be determined that a master equation approach is not feasible for that particular problem. It is possible also that a non-convergent series of terms leads also to a poor convergence of the numerical stochastic average of the SLN equation (\ref{LVN}). If the relative weight is smaller, then the equation (\ref{masteropen2}) can in principle be truncated and the master equation approach considered to be feasible. In that case, a truncated version of the evolution equation (\ref{masteropen2}) can be used, and it gives the non-approximated reduced density matrix of the system up to an error related to the truncation. 

We note that the procedure here used, can also be used by considering as a starting point instead of (\ref{LVN}), an SSE of the form \cite{alonso2005,alonso2007} 
\begin{eqnarray}
&&\frac{\partial G(z_i^* z_{i+1}|t_it_{i+1})}{\partial t_i}=
\big(-i H_S+L z^*_{i,t_i}-L^\dagger z_{i+1,t_i}\big)\cr
&\times&G(z_i^* z_{i+1}|t_it_{i+1})-L^\dagger \int_{t_{i+1}}^{t_i} d\tau 
\alpha(t_i-\tau)\cr
&\times&\frac{\partial}{\partial z^*_i}G(z_i^* z_{i+1}|t_it_{i+1}),
\label{chapuno323}
\end{eqnarray}
where $G(z^*_i z_{i+1} |t_i t_{i+1})=\langle z_{i}\mid {{\cal U}}_I (t_i,t_{i+1}) \mid z_{i+1} \rangle$
is the reduced propagator of the system. 
Here $ \mid z_{i+1} \rangle$ represents the initial state of the environment at time $t_{i+1}$, and $\mid z_{i}\rangle$ its final state at time $t_{i}$. The former is the generalization to an arbitrary initial state of the stochastic Schr{\"o}dinger equation derived in \cite{diosi1998,strunz1999} for an environment initially at zero temperature, i.e. $z_{i+1}=0$. Using (\ref{chapuno323}) opens the possibility to build the exact structure of master equations for problems where the coupling operator is non-Hermitian, the OQS is composed of many particles, or a more general initial state is considered. Finally, recently \cite{suess2014} a hierarchy of stochastic equations based on the SSE derived in \cite{diosi1998,strunz1999} have been derived. These are equations for successive partial derivatives of the wave-vector with respect to the noise, and have been proven to be particularly suitable to deal with exponential correlation functions or combinations of exponential correlation functions.
For exponential correlation functions, a hierarchy structure of master equations for fermionic environments has been derived in \cite{haertle2013}. A connection between these last two derivations and the one here proposed is an interesting open problem that may deserve further analysis. 

\textit{Acknowledgments}
The author gratefully acknowledges D. Alonso, and U. Schollw{\"o}ck for interesting discussions, and D. Alonso, J.I. Cirac, A. Ekert
and U. Schollw{\"o}ck for encouragement and support. This project was finantially supported by the Nanosystems Initiative Munich (NIM) (project No. 862050-2) and partially by the Spanish MICINN (Grant No. FIS2010-19998).

\section{Appendix: Test of the consistency condition}
Let us test the consistency condition with a very simple example. Let us assume the following evolution equation for the stochastic projector,
\bea
\frac{dP(t)}{dt}=\sum_\lambda L_\lambda(t)z_\lambda P(t),
\label{eq1}
\eea
where $L_\lambda(t)=g_\lambda(t)L(t)$, and $L(t)$ is a matrix within the system's Hilbert space that does not depend on any noise quantity $z_\lambda$. A partial derivation of the former equation with respect to $z_{\lambda'}$ is given as
\bea
\frac{\partial}{\partial z_{\lambda'}}\frac{dP(t)}{dt}=L_{\lambda'}(t)P(t)+\sum_\lambda L_\lambda(t)z_\lambda \frac{\partial P(t)}{\partial z_{\lambda'}}.
\label{eq2}
\eea
Integrating now (\ref{eq1}), we find
\bea
P(t)=P(0)+\int_0^t ds \sum_\lambda L_\lambda(s)z_\lambda P(s),
\eea
and deriving this solution with respect to $z_{\lambda'}$, we find
\bea
\frac{\partial P(t)}{\partial z_{\lambda'}}=\int_0^t ds L_{\lambda'}(s) P(s)+\int_0^t ds \sum_\lambda L_\lambda(s)z_\lambda \frac{\partial P(s)}{\partial z_{\lambda'}}.\nonumber
\eea
This equation, once derived with respect to $t$, gives rise to the same result as (\ref{eq2}), what proves that the time derivative and the partial derivative with respect to $z_\lambda$ commute, and therefore can be safely interchanged. 

\bibliography{/Users/ines.vega/Dropbox/Bibtexelesdrop}
\bibliographystyle{unsrt}

\end{document}